\documentclass[12pt,a4paper]{article}
\usepackage{epsfig}
\begin{document}
\title{ Signals of the littlest Higgs model with T-parity at  $e\gamma$ and $ep$ collisions}
\author{Chong-Xing Yue$^1$, Li Li$^2$, Shuo Yang$^1$, and Li-Na Wang$^1$\\
{\small $^1$ Department of Physics, Liaoning  Normal University,
Dalian 116029, P. R. China}
\\
{\small $^2$ Department of Mathematics, Liaoning  Normal University,
 Dalian 116029, P. R. China}
\\}
\date{\today}

\maketitle
\begin{abstract}
Littlest Higgs model with T-parity predicts the existence of the
neutral, weakly interacting, new gauge boson $B_{H}$, which can be
seen as an attractive dark matter candidate. We study production of
the new gauge boson $B_{H}$ via $e\gamma$ and $ep$ collisions. We
find that  $B_{H}$ can be abundantly produced via the subprocesses
$e^{-}\gamma\rightarrow L^{-}B_{H}$ and $\gamma q\rightarrow
B_{H}Q$, which might give rise to characteristic signals. Some
discussions about the $SM$ backgrounds for this kind of signals are
also given.

\vspace{1cm}

\end{abstract}

\newpage
\noindent{\bf I. Introduction}

Little Higgs models are proposed as an alternative solution to the
hierarchy problem of the standard model$(SM)$, which provide a
possible kind of electroweak symmetry breaking $(EWSB)$ mechanism
accomplished by a naturally light Higgs sector [1] (for review, see
[2]). In these models, the Higgs boson is a pseudo-Goldstone boson
and its mass is protected by a global symmetry and quadratic
divergence cancellations are due to contributions from new particles
with the same spin as the $SM$ particles. The dynamics of the Higgs
boson is described by a non-linear sigma model, valid up to the
cutoff scale $\Lambda\sim 4\pi f\sim 10 TeV$. The little Higgs
models generally predict the existence of the new gauge bosons,
fermions, and scalar particles at the $TeV$ scale. Some of these new
particles can generate characteristic signatures at the present  and
future collider experiments [3,4].

So far, a number of specific models have been proposed, which differ
in the assumed higher symmetry and in the representations of the
scaler multiplets. Among of these models, the littlest Higgs $(LH)$
model [1] is one of the simplest and phenomenologically viable
models, which has all essential features of the little Higgs theory.
However, the electroweak precision data produce rather severe
constraints on the free parameters of the $LH$ model, due to the
large corrections to low-energy observables from the new particles
and the triplet scalar vacuum expectation value $(VEV)$ [5]. To
alleviate this difficulty, a $Z_{2}$ discrete symmetry, named
'T-parity', is introduced into the $LH$ model, which is called $LHT$
model [6,7]. In the $LHT$ model, all the $SM$ particles are assigned
with an even T-parity, while all the new particles are assigned with
an odd T-parity, except for the little Higgs partner of the top
quark. In the $LHT$ model, the T-parity is an exact symmetry, the
$SM$ gauge bosons (T-even) do not mix with the T-odd new gauge
bosons, and thus the electroweak observables are not modified at
tree level. Beyond the tree level, small radiative corrections
induced by the $LHT$ model to the electroweak observables still
allow the scale parameter $f$ to be lower than 1 $TeV$ [6,8]. Since
the contributions of the little Higgs models to observables are
generally proportional to the factor $1/f^{2}$, so a lower $f$ is
very important for the phenomenology of these models and might
produce rich signatures at the collider experiments
[9,10,11,12,13,14,15].

An interesting feature of the $LHT$ model is that it predicts the
existence of the lightest T-odd particle, which is stable,
electrically neutral, and weakly interacting, new gauge boson
$B_{H}$. It has been shown that it can be seen as an attractive dark
matter candidate and might generate observable signals at the
hadronic colliders [11,15]. In this paper, we will discuss the
production of the new gauge boson $B_{H}$ via $e\gamma$ and $ep$
collisions and see whether it can produce observable signatures in
the future high energy $e^{+}e^{-}$ and $ep$ collider experiments,
i.e. $ILC$ [16] and $THERA$ [17].

In the rest of this paper, we will give  our results in detail. In
section II, we give some free parameters, which are related our
calculation. The production of the gauge boson $B_{H}$ via the
processes $e^{+}e^{-}\rightarrow e^{-}\gamma\rightarrow L^{-}B_{H}$
and $ep\rightarrow \gamma q\rightarrow B_{H}Q$ at the $ILC$ and
$THERA$ experiments are considered in section III and section IV,
respectively. The relevant phenomenology discussions are given in
these sections. Our conclusions are given in section V.

 \noindent{\bf II. The relevant parameters about our calculation }

Similar with the $LH$ model, the $LHT$ model [6,7] is based  on an
$SU(5)/SO(5)$ global symmetry breaking pattern and the Higgs doublet
of the $SM$ is identified with a subset of the Goldstone boson
fields associated with this breaking. A subgroup $[SU(2)_{1}\times
U(1)_{1}]\times[SU(2)_{2}\times U(1)_{2}]$ of the $SU(5)$ is gauged,
and at the scale $f$ it is broken into the $SM$ electroweak symmetry
$SU(2)_{L}\times U(1)_{Y}$. This breaking scenario gives rise to
four new gauge bosons $W_{H}^{\pm}$, $Z_{H}$, and $B_{H}$. However,
in the $LHT$ model, T-parity is an automorphism which exchanges the
$[SU(2)_{1}\times U(1)_{1}]$ and $[SU(2)_{2}\times U(1)_{2}]$ gauge
symmetries. Under this transformation, the $SM$ gauge bosons
$W^{\pm}$, $Z$, and $\gamma$ are T-even and the new gauge bosons
$W_{H}^{\pm},Z_{H}$, and $B_{H}$ are T-odd.

Among these new particles, the neutral gauge boson $B_{H}$ is the
lightest particle which can be seen as an attractive dark matter
candidate [9,10,11,12]. At the order of $\nu^{2}/f^{2}$, the mass of
the neutral gauge boson $B_{H}$ can be approximately written as:
\begin{equation}
M_{B_{H}}\simeq\frac{g'f}{\sqrt{5}} [1-\frac{5\nu^{2}}{8f^{2}}],
\end{equation}
where $g'$ is the $SM$ $U(1)_{Y}$ gauge coupling constant, and
$\nu\simeq246GeV$ is the electroweak scale. At the order of
$\nu^{2}/f^{2}$, the couplings of the neutral gauge boson $B_{H}$
with the first or second family fermions and their corresponding
little Higgs partners can be approximately written as [10]:
\begin{equation}
\pounds=iYg'\bar{Q}\gamma_{\mu}p_{L}qB^{\mu}_{H}+h.c.,
\end{equation}
where $Y=1/10$, $q=u,d,c,s,e,$ or $\mu$,  and $Q$ is the T-odd
partner of the T-even $SM$ fermion $q$. The more exact expressions
of these couplings have been recently given in Ref.[15]. According
the formula given in Ref.[15], the coupling constant $Y$ is
generally not equal to $1/10$. However, for the scale parameter $
f\geq 500GeV $, the coupling constant $Y$ is very approach to this
value. Thus, as numerical estimation, we will assume that there are
a universal coupling constant  $Y=1/10$ for the couplings of the
gauge boson $B_{H}$ to the $SM$ quark and lepton partners, as shown
in Eq.(2). Furthermore, as doing in Refs. [10,11], we also assume a
universal T-odd fermion mass $M_{Q}$ for the little Higgs partners
of the first and second family fermions, and take $M_{Q}\simeq
M_{B_{H}}+20GeV$.

In the $LHT$ model, because of  the T-parity, the $SM$ gauge bosons
(T-even) do not mix with the T-odd new gauge bosons. Thus, the
electroweak  observables are not modified at tree-level. The new
heavy T-odd particles, such as T-odd gauge bosons, T-odd fermions,
and T-odd triplet scalars, can only has contributions to the
electroweak observables at loop level, which are typically small. So
the scale parameter $f$ can be as low as $500GeV$ [8]. In this
paper, we will take the scale $f$ as free parameter and assume that
its value is in the range of $500GeV\sim2000GeV$.

\noindent{\bf III. Production of the new gauge boson $B_{H}$ at
$e\gamma$ collision}

There are two Feynman diagrams, depicted in Fig.1, contributing
production of the new gauge boson $B_{H}$ associated with the T-odd
partner $L^{-}$ of the lepton $e^{-}$ via $e^{-}\gamma$ collision.
The corresponding scattering amplitude can be written as:
\begin{equation}
M=eYg'Q_{L} \bar{u}(L)[\frac{\not\varepsilon_{2} P_{L} (\not
P_{\gamma}+\not P_{e}+m_{e}) \not\varepsilon_{1}} {\hat
s-m_{e}^{2}}+\frac{\not\varepsilon_{1}(\not P_{L}-\not
P_{\gamma}+M_{L})\not\varepsilon_{2}P_{L}} {\hat u-M_{L}^{2}}]u(e),
\end{equation}
where $\hat s=(P_{\gamma}+P_{e})^{2}=(P_{B_{H}}+P_{L})^{2}$, $\hat
u=(P_{L}-P_{\gamma})^{2}=(P_{e}-P_{B_{H}})^{2}$. $\varepsilon_{1}$
and $\varepsilon_{2}$ are the polarization vectors of the gauge
bosons $\gamma$ and $B_{H}$, respectively. $M_{L}$ is the mass of
the $SM$ lepton partner $L^{-}$ and  taken as $M_{L}=M_{Q}=
M_{B_{H}}+20GeV$.

The effective cross section $\sigma_{1}(s)$ at a $ILC$ with the
center-of-mass energy $\sqrt{s}$ can be obtained by folding the
cross section $\hat{\sigma}_{1}(\hat{s})$ for the subprocess
$e^{-}\gamma\rightarrow L^{-}B_{H}$ with the photon distribution
function $f_{\gamma/e}$ [18]:
\begin{equation}
\sigma_{1}(s)=\int^{0.83}_{(M_{B_{H}}+M_{L})^{2}/s}
dx\hat{\sigma}_{1}(\hat{s})f_{\gamma/e}(x),
\end{equation}
where $\hat{s}=xs$.
\begin{figure}[htb] \vspace{-6cm}
\hspace{-2cm} \epsfig{file=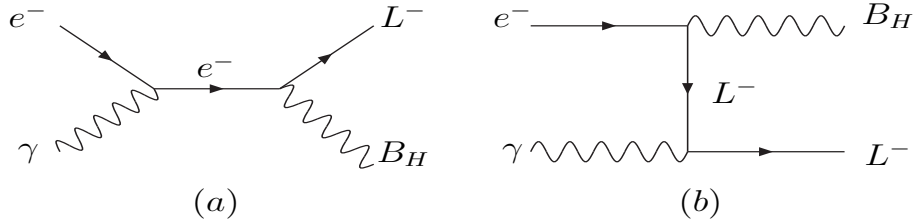,width=700pt,height=800pt}
\vspace{-20cm}
 \caption{Feynman diagrams contribute to the process $e^{-}\gamma \rightarrow L^{-}B_{H}$. }
 \label{ee}
\end{figure}

From above equations, we can see that, except the $SM$ input
parameters, the effective cross section $\sigma_{1}(s)$ is dependent
on the scale parameter $f$ and the center-of-mass energy $\sqrt{s}$.
So, we plot the cross section $\sigma_{1}(s)$ as a function of $f$
for $\sqrt{s}=1TeV$ in Fig.2. One can see from Fig.2 that the value
of $\sigma_{1}(s)$ is strongly dependent on the scale parameter $f$.
For $500GeV\leq f \leq2000GeV$, the value of the production cross
section $\sigma_{1}$ is in the range of $16.4fb\sim 1.5fb$. If we
assume that the $ILC$ experiment with $\sqrt{s}=1TeV$ has a yearly
integrated luminosity of $\pounds=500fb^{-1}$, then there will be
several hundreds up to ten thousands of the new gauge boson $B_{H}$
associated with the $SM$ lepton partner $L^{-}$ to be generated per
year.

For $M_{L}=M_{B_{H}}+20GeV$, the lepton partner $L$ mainly decays to
$B_{H}l$ and there is $Br(L\rightarrow B_{H}l)\simeq 1$ [10,11].
Thus, the signal of the process $e^{-}\gamma\rightarrow L^{-}B_{H}$
should be an isolated charged lepton associated with large missing
energy. The most large backgrounds  for the signal $l^{-}+\not E$
come from the $SM$ processes $e^{-}\gamma\rightarrow
e^{-}Z\rightarrow e^{-}\nu\nu$ and $e^{-}\gamma\rightarrow
W^{-}\nu_{e}\rightarrow l^{-}\nu\nu_{e}$. The scattered electron in
the process $e^{-}\gamma\rightarrow e^{-}Z$ has almost the same
energy $E_{e}=\sqrt{s}/2$ for $\sqrt{s}\gg M_{Z}$. Thus, the process
$e^{-}\gamma\rightarrow e^{-}Z$ could be easily distinguished from
the signal [19]. So, the most serious background process is
$e^{-}\gamma\rightarrow W^{-}\nu_{e}\rightarrow l^{-}\nu\nu_{e}$. To
discuss whether the possible signals of the $LHT$ model can be
detected via the process $e^{-}\gamma\rightarrow L^{-}B_{H}$ in the
future $ILC$ experiments, we future calculate the ratio of the
signal over the square root of the background
$(R_{1}=N_{1}/\sqrt{B})$. Our numerical results show that the $SM$
backgrounds are much large and the value of the ratio $R_{1}$ is
smaller than $0.5$ in most of the parameter space of the $LHT$
model. Thus, the possible signals of the $LHT$ model are very
difficult to be detected via $e\gamma$ collision in the future $ILC$
experiment with $\sqrt{s}=1TeV$ and $\pounds=500fb^{-1}$.

\begin{figure}[htb] \vspace{-0.5cm}
\begin{center}
\epsfig{file=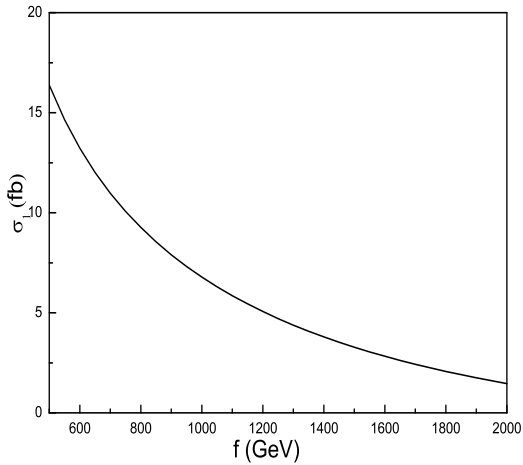,width=220pt,height=205pt}
\hspace{-0.5cm}\vspace{-1.0cm}
 \caption{The effective cross section $\sigma_{1}$ as a function of the scale parameter $f$ for the
\hspace*{1.9cm} center-of-mass energy $\sqrt{s}=1TeV$.}
 \label{ee}
\end{center}
\end{figure}

It is well known that a appropriate cut on the $SM$ background can
generally enhance the ratio of signal over square root of the
background. It has been shown that, with the suitably cut on the
final lepton transverse momentum and rapidity, the $SM$ background
$l^{-}\nu\nu_{e}$ can be reduced by more than one order of magnitude
$[19]$. Furthermore, beam polarization of the electron and positron
beams would lead to a substantial enhancement of the production
cross sections for some specific processes with a suitably chosen
polarization configuration. Thus, we expect that we might use these
methods to discriminate the signal $l^{-}+\not E$ from the $SM$
background.

\noindent{\bf IV. Production of the neutral gauge boson $B_{H}$ at
$ep$ collision}

Similar to the process $e^{-}\gamma\rightarrow L^{-}B_{H}$,
production of the new gauge boson $B_{H}$ associated with the T-odd
partner of the $SM$ quark at the $THERA$ proceeds via the s-channel
and t-channel Feynman diagrams, as shown in Fig.3. The invariant
scattering amplitude for the subprocess $\gamma q\rightarrow
B_{H}Q(q=u,c,d$ or $s)$ can be written as: \vspace{-0.5cm}

\begin{equation}
M=eYg'Q_{Q} \bar{u}(L)[\frac{\not\varepsilon_{2} P_{Q} (\not
P_{\gamma}+\not P_{q}+m_{q}) \not\varepsilon_{1}} {\hat
s-m_{q}^{2}}+\frac{\not\varepsilon_{1}(\not P_{Q}-\not
P_{\gamma}+M_{Q})\not\varepsilon_{2}P_{Q}} {\hat u-M_{Q}^{2}}]u(q),
\end{equation}
where $\hat s=(P_{\gamma}+P_{q})^{2}=(P_{B_{H}}+P_{Q})^{2}$, $\hat
u=(P_{Q}-P_{\gamma})^{2}=(P_{q}-P_{B_{H}})^{2}$. $\varepsilon_{1}$
and $\varepsilon_{2}$ are the  polarization vectors of the gauge
bosons $\gamma$ and $B_{H}$, respectively.

\begin{figure}[htb] \vspace{-6cm}
\hspace{-5cm} \epsfig{file=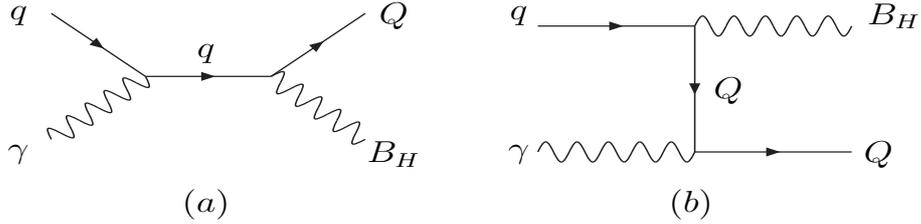,width=700pt,height=800pt}
\vspace{-20cm}
 \caption{Feynman diagrams contribute to the process $\gamma q \rightarrow B_{H}Q$ }
 \label{ee}
\end{figure}

After calculating the cross section $\hat{\sigma}_{i}(\hat{s})$ of
the subprocess $\gamma q\rightarrow B_{H}Q$, the total cross section
$\sigma_{2}(s)$ of $B_{H}Q$ production can be obtained by folding
$\hat{\sigma}_{i}(\hat{s})$ with the parton distribution functions:
\begin{equation}
\sigma_{2}(s)=\sum_{i}\int^{0.83}_{\tau_{min}}d\tau\int^{1}_{\tau/0.83}
\frac{dx}{x}f_{\gamma/e}(\frac{\tau}{x})f_{i/p}(x)\hat{\sigma_{i}}(\hat{s}),
\end{equation}
with $\hat{s}=\tau s$, $\tau_{min}=\frac{(M_{B_{H}}+M_{Q})^{2}}{s}$,
$i=u,c,d$ and $s$. The backscattered high energy photon distribution
function has been given in Ref.[18]. In our calculation, we will
take the $CTEQ6L$ parton distribution function for $f_{i/p}(x)$
[20].

Our numerical results are shown in Fig.4, in which we have taken
$\sqrt{s}=1TeV$. One can see from Fig.4 that the production cross
section $\sigma_{2}$ is  smaller than $\sigma_{1}$ in all of the
parameter space. This is because, compared with $\sigma_{1}$,
$\sigma_{2}$ is suppressed by the parton distribution function
$f_{i/p}(x)$ and the charge factor $Q_{q}^{2}$. For $500GeV\leq
f\leq 2000 GeV$, the value of the cross section $\sigma_{2}$ is in
the range of $14.6fb \sim 2.4\times10^{-3} fb.$ If we assume that
the $THERA$ collider with $\sqrt{s}=1TeV$ has a yearly integrated
luminosity of $\pounds=470fb^{-1}$[17], then there will be several
 and up to hundreds of the $B_{H}Q$ events to be generated
per year.
\begin{figure}[htb]\vspace{-0.5cm}
\begin{center}
\epsfig{file=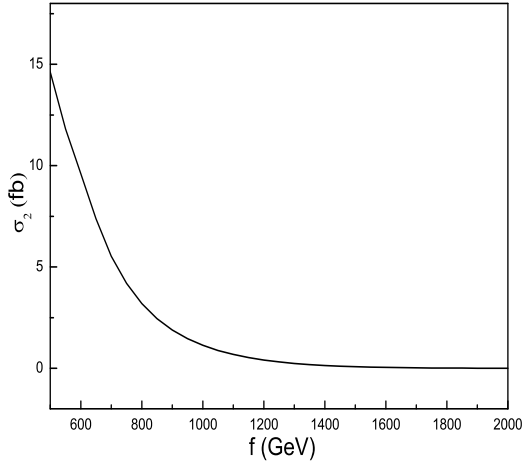,width=220pt,height=205pt}
\hspace{-0.5cm}\vspace{-1cm}
 \caption{The effective cross section $\sigma_{2}$ as a function of the scale parameter $f$ for the
\hspace*{1.9cm} center-of-mass energy $\sqrt{s}=1TeV$.}
 \label{ee}
 \end{center}
\end{figure}

From above discussions, we can see that the $SM$ quark partner $Q$
mainly decays to $B_{H}q$ $(q=u,c,d$ or $s)$ and there is
$Br(Q\rightarrow B_{H}q)\simeq 1$. In this case, the signal of the
$B_{H}Q$ production at the $THERA$ collider is one jet plus large
missing energy. The backgrounds of the one jet$+\not E$ signal
mainly come from the $SM$ charged current(CC) process $e
p\rightarrow \nu X$ and the SM subprocess $\gamma q\rightarrow Zq$
with $Z\rightarrow \nu\bar{\nu}$. Measurement and QCD analysis of
the production cross section of the $SM$ CC process $e p\rightarrow
\nu X$ at the HERA  collider has been extensively studied [21]. Its
cross section is very larger than that of the process $e p
\rightarrow \gamma q\rightarrow Zq$. Thus, we only take the $SM$ CC
process $e p\rightarrow \nu X$ as background of the process $e p
\rightarrow \gamma q\rightarrow B_{H} Q$. We have checked the $SM$
background and found that it is well above the signal of the one jet
plus missing energy from the $LHT$ model. However, this  kind of the
$SM$ backgrounds have been well studied and will be precisely
measured at the $THERA$ collider experiments, one can still look for
excess in the one jet$+\not E$ signal to search for the possible
signals of the $LHT$ model.

\noindent{\bf V. Conclusions}

To avoid the severe constraints from the electroweak precision data
on the $LH$ model, the T-parity is introduced into this model, which
forms the $LHT$ model. A interesting feature of the $LHT$ model is
that it predicts the existence of the neutral, weakly interacting,
new gauge boson $B_{H}$, which can be seen as an attractive dark
matter candidate. This model can generate vary different signals
from those for the $LH$ model in the present or future high energy
experiments.

In this paper, we discuss production of the new gauge boson $B_{H}$
predicted by the $LHT$ model at the $ILC$ and $THERA$ collider
experiments via considering the subprocesses $e^{-}\gamma\rightarrow
L^{-}B_{H}$ and $\gamma q\rightarrow B_{H}Q$. We find that the new
gauge boson $B_{H}$ can be abundantly produced at these collider
experiments. The signals of the associated production of
$L^{-}B_{H}$ and $B_{H}Q$ are an isolated charged lepton with large
missing energy and one jet with large missing energy, respectively.
Thus, the possible signals of the $LHT$ model might be detected at
the $ILC$ and $THERA$ experiments by searching for one jet (or
charged lepton) with large missing energy. We further give some
discussions about the $SM$ backgrounds for this kind of signals.
Despite the fact that the $SM$ backgrounds are much large, it is
also need to careful study the $SM$ backgrounds in order to search
for these signals of the $LHT$ model in the future $ILC$ and $THERA$
collider experiments.

Certainly, if the new gauge boson $B_{H}$ and the $SM$ quark partner
$Q$ are enough light, the process $e p \rightarrow \gamma
q\rightarrow B_{H} Q$ can occur at the HERA with $ \sqrt{s}= 320
GeV$, which can also generate the characteristic signals at the HERA
experiments. However, considering the constraints on the mass
parameters $M_{B_{H}} $ and  $M_{Q} $, we have not calculated the
cross section of this process at the HERA collider experiments.

Other specific models beyond the $SM$, such as $SUSY$ and extra
dimension models, can also generate the similar signals with those
of the $L^{-}B_{H}$ and $B_{H}Q$ associated production. More
studying about this kind of signals is needed and it will be helpful
to discriminate various specific models beyond the $SM$ in the
future high energy experiments.

 \vspace{0.5cm} \noindent{\bf Acknowledgments}

We would like to thank Prof. C. Diaconu for valuable helping and
comments on our work. This work was supported in part by Program for
New Century Excellent Talents in University(NCET-04-0290), the
National Natural Science Foundation of China under the Grants
No.10475037 and 10675057.


\vspace{1.0cm}


\newpage
\begin{thebibliography}{99}

\bibitem{1}
        N. Arkani-Hamed, A. G. Cohen, E. Katz, A. E. Nelson, {\em JHEP}
        {\bf 0207}, 034(2002).
\bibitem{2}
        M. Schmaltz and D.Tucker-Smith, {\em Ann. Rev. Nucl. Parti. Sci.} {\bf
         55}, 229(2005).
\bibitem{3}
        T. Han, H. E. Logan, and L. T. Wang, {\em JHEP} {\bf 0601}, 099(2006).
\bibitem{4}
        M. Perelstein, {\em Prog.Part.Nucl.Phys.} {\bf
         58}, 247(2007).
\bibitem{5}
        J. L. Hewett, F. J. Petriello and T. G. Rizzo, {\em JHEP} {\bf
        0310}, 062(2003). C. Csaki, J.Hubisz, G. D. Kribs, P. Meade, and J.
        Terning, {\it Phys. Rev. D}{\bf 67}, 115002(2003);
        C. Csaki et al.,  {\it Phys. Rev. D}{\bf 68}, 035009(2003); R. Casalbuoni,
        A. Deandrea, M. Oertel, {\em JHEP} {\bf 0402}, 032(2004);
         Mu-Chun Chen and S. Dawson, {\it Phys. Rev. D}{\bf70}, 015003(2004);
          Chong-Xing Yue and Wei Wang, {\it Nucl. Phys. B}{\bf
         683}, 48(2004);  W. Kilian and J. Reuter, {\it Phys. Rev. D}{\bf
          70}, 015004(2004);  T. Gregoire,
        D. R. Smith and J. G. Wacker, {\it Phys. Rev. D}{\bf
        69}, 115008(2004); Mu-Chun Chen, {\em Mod. Phys. Lett.
        A}{\bf21}, 621(2006).
\bibitem{6}
        H. C. Cheng and I. Low, {\em JHEP} {\bf 0309}, 051(2003); {\em JHEP} {\bf 0408}, 061(2004) .
\bibitem{7}
        I. Low, {\em JHEP} {\bf 0410}, 067(2004).
\bibitem{8}
        J. Hubisz, P. Meade, A. Noble and M. Perelstein, {\em JHEP} {\bf 0601}, 135(2006).
\bibitem{9}
        J. Hubisz, P. Meade, {\it Phys. Rev. D}{\bf 71}, 035016(2005).
\bibitem{10}
         A. Birkedal, A. Noble, M.Perelstein, and A. Spary, {\it Phys. Rev. D}{\bf 74}, 035002(2006).
\bibitem{11}
        C. S. Chen, Kingman Cheung, and T.-C. Yuan, {\it hep-ph}/{\bf 0605314}.
\bibitem{12}
        M. Asano et al., {\it hep-ph}/{\bf 0602157}; A. Martin, {\it hep-ph}/{\bf 0602206}.
\bibitem{13}
        J. Hubisz, S. J. Lee, and G. Paz, {\em JHEP} {\bf 0606}, 041(2006);
        M. Blanke et al., {\it hep-ph}/{\bf 0605214}; A.
        J. Buras, A. Poschenrieder, S. Uhlig, W. A. Bardeen, {\it hep-ph}/{\bf
        0607189}; A. Freitas and D.Wyler, {\it hep-ph}/{\bf
        0609103}.
\bibitem{14}
        C. O. Dib, R. Rosenfeld, A. Zerwekh, {\em JHEP} {\bf
        0605}, 074(2006); H. C. Cheng, I. Low, and L. T. Wang, {\it Phys. Rev. D}{\bf 74}, 055001(2006);
         C. R. Chen, K. Tobe and C.-P. Yuan, {\em Phys. Lett.
        B}{\bf640}: 263(2006).

\bibitem{15}
       A. Belyaev, Chuan-Ren Chen, K. Tobe,
        C.-P. Yuan, {\it hep-ph}/{\bf 0609179}.
\bibitem{16}
           T. Abe et al. [American Linear Collider Group], {\em hep-ex}/{\bf0106057};
        J. A. Aguilar-Saavedra et al.
        [{\it ECFA/DESY LC} Physics Working Group], {\em hep-ph}/{\bf
        0106315}; K. Abe et al. [{\em ACFA} Linear Collider Working Group],
        {\em hep-ph}/{\bf 0109166}; G. Laow et al.,
        {\em ILC} Techinical Review Committee, second report, 2003,
        SLAC-R-606.

\bibitem{17}
           H. Abramowicz et al.({\it TESLA-N Study Group Colloboration}), {\it Technical Design Report},
         DESY 01-011 preprint; A. K. Cifici, S. Su Handoy, \"{O} Yavas, {\it in Proc. of EPAC2000}
        P.388; M. Klein, {\it THERA-electron-proton scattering at
        $\sqrt{s}=1TeV$}, talk given at DIS'2000, Liverpool, April
        25-30, 2000.
        P. J. Bussey, {\it Int. J. Mod. Phys. A}{\bf17}, 1065(2002).
\bibitem{18}
        I. F. Ginzbury et al., {\em  Nucl. Instrum. Meth. Phys. Res. Sec A}, {\bf 21}, 5(1984);
        V. I. Telnov, {\em  Nucl. Instrum. Meth. Phys. Res. Sec A}, {\bf 294}, 72(1990).
\bibitem{19}
        F. M. Renard, {\em Z. Phys. C.}{\bf 14}, 209(1982); F. Cuypers, G. J. Van Oldenborgh, and R. Ruckl,
        {\it Nucl. Phys. B}{\bf 383}, 45(1992); M. Raidal, {\it Nucl. Phys. B}{\bf
         441}, 49(1995); E. M. Gregores, M. C. Gonzalez-Garcia, and S.
         F. Novaes, {\it Phys. Rev. D}{\bf 56}, 2920(1997).
\bibitem{20}
        J. Pumplin et al., {\em JHEP} {\bf 0207}, 012(2002); D. Stump et
        al., {\em JHEP} {\bf 0310}, 046(2003).

\bibitem{21}
        C. Adloff et al.[H1 Collaboration], { \it Eur. Phys. J. C}{\bf 30}, 1(2003); S. Chekanov et al.
        [ZEUS Collaboration], { \it Eur. Phys. J. C}{\bf 32}, 1(2003);  A. Aktas et al.[H1 Collaboration],
          { \it Phys. Lett. B}{\bf 634}, 173(2006).
\end{thebibliography}
\end{document}